\documentclass[prb,aps,twocolumn,amsmath,amssymb,floatfix,
superscriptaddress]{revtex4}
\usepackage[dvips]{graphics}
\usepackage{color}
\definecolor{dred}{rgb}{0,0,0.6}
\usepackage{graphicx}  
\usepackage{dcolumn}   
\usepackage{bm}        
\usepackage{amssymb}   
\usepackage{amsmath}
\usepackage{braket}
\usepackage[mathscr]{euscript}
\usepackage{color}
\usepackage{hyperref}
\begin{document}
\title{A driven fractal network: Possible route to efficient thermoelectric application}

\author{Kallol Mondal}
\email[E-mail: ]{kallolsankarmondal@gmail.com}
\affiliation{Physics and Applied Mathematics Unit, Indian Statistical Institute, 203 Barrackpore Trunk Road, Kolkata-700108, India}

\author{Sudin Ganguly}
\email[E-mail: ]{sudinganguly@gmail.com}
\affiliation{Physics and Applied Mathematics Unit, Indian Statistical Institute, 203 Barrackpore Trunk Road, Kolkata-700108, India}

\author{Santanu K. Maiti}
\email[E-mail: ]{santanu.maiti@isical.ac.in}
\affiliation{Physics and Applied Mathematics Unit, Indian Statistical Institute, 203 Barrackpore Trunk Road, Kolkata-700108, India}


\begin{abstract}
An essential attribute of many fractal structures is self-similarity.  A Sierpinski gasket (SPG) triangle is a promising example of a fractal lattice that exhibits localized energy eigenstates.  In the present work, for the first time we establish that a mixture of both extended and localized energy eigenstates can be generated yeilding  mobility edges at multiple energies in presence of a time-periodic driving field.  We obtain several compelling features by studying the transmission and energy eigenvalue spectra. As a possible application of our new findings, different thermoelectric properties are discussed, such as  electrical conductance, thermopower, thermal conductance due to electrons and phonons. We show that our proposed method indeed exhibits highly favorable thermoelectric performance. The time-periodic driving field is assumed through an arbitrarily polarized light, and its effect is incorporated via Floquet-Bloch ansatz. All transport phenomena are worked out using Green's function formalism following the Landauer-B\"{u}ttiker prescription.  
\end{abstract}

\maketitle

\section{Introduction}

Deterministic fractals are neither a perfectly ordered nor a completely disordered structure but somewhat in between them. Unlike the Anderson localization~\cite{anderson}, in deterministic fractals, the localization occurs due to the finite ramifications and self-similar structures such as Sierpinski gasket (SPG). The existence of highly degenerate localized states and the Cantor set energy spectrum are the hallmarks of the SPG structures~\cite{domany}. These localized states become delocalized in the presence of a magnetic field~\cite{banavar}. Owing to such distinctive properties, SPG structures have been studied extensively over the years in many contexts, and several other unique features have been observed~\cite{rammal,gordon,wasch,wang1,wang2,mayer,maiti-prb-spg,veen1,veen2}.

Recently, it has been suggested that a spatial anisotropy can be tuned using a time-periodic driving field~\cite{gomez-prl} and thus, it is possible to manipulate the material and topological properties~\cite{runder}. Such a possibility led us to think about the structure-induced localization phenomenon in SPG structures. What will be the nature of the energy eigenstates in the presence of a driving field is still an open question to the best of our concern. We are particularly interested in finding a mobility edge that separates localized energy eigenstates from the extended ones.

Usually, in 1D and 2D systems, localization to delocalization transition does not occur in the presence of uncorrelated disorder since all the states are localized~\cite{anderson,lee}. However, quasi-periodic lattices, such as one-dimensional (1D) Aubry-Andr\'{e} (AA) chains\cite{aa} exhibit a delocalization to localization transition at a critical disorder strength. Below this critical value, all states are extended, and beyond that, all the states become localized. For such a situation, mobility edge does not  occur as the mixture of extended and localized states is no longer available.  However, mobility edge has been observed in 1D AA chains~\cite{sds-prl-2010} and ladder networks~\cite{sil-prl} in the presence of higher-order hopping integral(s). So far, no attempt has been made to detect mobility edge in fractal lattices, which essentially motivates us to probe into it. It is well known that all the states of an SPG lattice become localized in the asymptotic limit due to the structure-induced localization.

As the mobility edge is directly associated with the metal-insulator transition, the electronic transmission function will be highly asymmetric around the mobility edge. The asymmetry in the transmission probability is the key requirement to get a favorable thermoelectric (TE) response~\cite{dubi}. Hence given the existence of mobility edge, SPG structure could be a potential candidate for the TE applications, and we investigate such a promising aspect in the present work. SPG structures already have been fabricated experimentally with several materials, such as submicrometer-width Al wires~\cite{gordon-prl}, and also very recently from aromatic compounds~\cite{js-natchem}, metal-organic compounds~\cite{li-acs}, and by manipulation on CO molecules of Cu(111) surface~\cite{snk}. Therefore, with the recent experimental realizations of SPG structures, we believe that our proposition can be substantiated in a suitable laboratory.

The SPG is driven by an arbitrarily polarized light, and its effect is incorporated through the standard Floquet-Bloch ansatz in the minimal coupling scheme~\cite{gomez-prl,sambe,grifoni,lght1,lght2, sudin-carbon, sudin-jap, maiti-prb2020}. The mobility edge is detected by superimposing the energy eigenvalues of the non-interacting electrons and the two-terminal transmission probability. We compute the later one by using the well-known Green's function formalism, based on Landauer-B\"{u}ttiker prescription~\cite{etms,qtat}. The TE performance is studied by evaluating the electrical conductance, thermopower, and thermal conductance due to electrons utilizing Landauer prescription~\cite{dubi,lam}. Since at finite temperature, the effect of phonons cannot be ignored, we also give an estimation of the thermal conductance due to phonons for a precise measurement of the TE efficiency employing the non-equilibrium Green's function formalism~\cite{zhang-ph,hopkins-prb,aghosh}.

The key findings of our work are: (i) generation of multiple mobility edges in the presence of a driving field, and (ii) achieving of high thermoelectric performance due to the existence of asymmetric transmission function around the mobility edges. Our analysis can be utilized to design efficient thermoelectric devices at the nanoscale level and to study some fascinating phenomena in similar kind of fractal lattices and other topological systems.

The rest of the work is organized as follows. In Sec. (2), we present our model Hamiltonian for the SPG system in the presence of an arbitrarily polarized light. In this section, we also present a brief theoretical description for the calculations of two-terminal transmission probability and different TE quantities, including the thermal conductance due to phonon. All the results are critically investigated in Sec. (3). Finally, in Sec. (4), we conclude our essential findings.


\section{SPG network and theoretical formulation}
\subsection{SPG and Hamiltonian}

\begin{figure}[t!]
\centering
\includegraphics[width=0.45\textwidth]{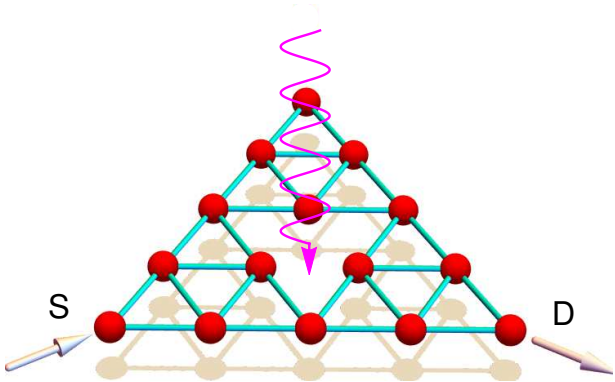} 
\caption{(Color online). Schematic view of an irradiated 3rd generation SPG fractal network. The atomic sites are located at the vertices of each equilateral triangle, as shown by red solid spheres. The SPG is attached to two electrodes (source $S$ and drain $D$). The electrodes are kept at two different temperatures $T + \Delta T/2$ and $T - \Delta T/2$, where $ \Delta T$ is infinitesimally small.}
\label{spg}
\end{figure}

The SPG network is perfectly self-similar with three non-overlapping copies of the previous generation, and every triangular plaquette is a replica of the full structure as shown in Fig.~\ref{spg} where a third-generation SPG is depicted schematically. To evaluate the transmission probabilities and investigate TE performance, we clamp the SPG network between to perfect, reflectionless, semi-infinite, and 1D electrodes, namely the source ($S$) and drain ($D$). A temperature difference $\Delta T$ is set among these electrodes $S$ and $D$, and this $\Delta T$ is chosen to be
small enough such that we can work in the linear response regime. An arbitrarily polarized light (magenta curve) is incident on the SPG perpendicular to the lattice plane while the electrodes are free from any kind of irradiation.

To describe the system, we use the tight-binding framework, which can potentially capture the essential physics of quantum transport. In this framework, the model Hamiltonian consists of four parts, as described below
\begin{eqnarray}
H & = &H_{\text{SPG}} + H_\text{S} + H_\text{D} + H_\text{C}
\end{eqnarray}
where $H_{\text{SPG}}$, $ H_\text{S(D)}$ and $H_\text{C}$ represent the sub-parts of the Hamiltonian associated with the SPG network, the source (drain), and the coupling between semi-infinite leads and the SPG network, respectively. The coupling part of the Hamiltonian consists of two terms; one is the coupling between the source and SPG network, and the other one is the coupling between the drain and SPG network. In the absence of light irradiation, these sub-Hamiltonians are expressed as follows.
\begin{subequations}
\begin{eqnarray}
H_{\text{SPG}} & = &\sum_n \epsilon_n c_n^\dagger c_n + \sum_{\langle nm \rangle} t_{nm}( c_n^\dagger c_m + \text{h.c.}) \\
H_{\rm S} &=& H_{\rm D} = \epsilon_0\sum\limits_{n} d_n^{\dagger} d_n +
t_0\sum\limits_{\langle nm\rangle}\left(d_n^{\dagger} d_m + h.c.\right),\\
H_{\rm C}  &=& H_{\rm S,\rm SPG} + H_{\rm D, \rm SPG} \nonumber \\
& = &\tau_S\left(c_p^{\dagger} d_0 + h.c.\right) + \tau_D\left(c_q^{\dagger} d_{N+1} + h.c.\right).
\end{eqnarray}
\label{ham-nolight}
\end{subequations}
The annihilation operators $c,d$ and their hermitian counterparts $c^\dagger,d^\dagger$ satisfy the usual fermionic commutation relations. $\left(c,c^\dagger\right)$ are associated with the SPG network and $\left(d,d^\dagger\right)$ with the source and drain. $\epsilon_n$ represents the on-site potential at the $n$-th site. $t_{nm}$ denotes the nearest-neighbor hopping (NNH) integral in the SPG in the absence of light. On-site potential $\epsilon_0$ and hopping amplitude $t_0$ are assumed to be the same for both the source and drain. The coupling strength between the source and SPG is $\tau_S$, and that between the drain and SPG is $\tau_{D}$. The source and drain are connected to the SPG at the $p$-th and $q$-th sites, respectively.

\subsection{Incorporation of light irradiation } When a system is irradiated with light, the system becomes a periodically driven one. Under this situation, the problem becomes quite complicated and challenging as well. But in the minimal coupling regime, such a time-dependent problem can be simplified using Floquet-Bloch ansatz~\cite{gomez-prl,sambe,grifoni,lght1,lght2}. Following the Floquet approximation, the effect of light incorporation can be taken care of through a vector potential $\mathbf{A}(\tau)$. In the tight-binding framework, the vector potential manifests itself in the hopping integral through Peierls substitution $\frac{e}{c\hslash}\int {\mathbf A}(\tau)\cdot d{\mathbf l}$, where the symbols $e$, $c$, and $\hslash$ carry their usual meaning. Without losing any generality, we can write the vector potential in the form $\mathbf{A(\tau)} = (A_x \sin(\Omega \tau), A_y \sin(\Omega \tau + \phi),0)$, which represents an arbitrarily polarized field in the X-Y plane. $A_x$ and $A_y$ are the field amplitudes, and $\phi$ is the phase. Depending upon the choices $A_x, A_y$, and $\phi$, we can get different polarized lights, such as circularly, linearly, or elliptically polarized lights. After rigorous mathematical calculation, the effective hopping integral in the presence of irradiation gets the form 
\begin{equation}
t_{nm} \rightarrow t_{nm}^{pq} = t_{nm} \times \frac{1}{\mathbb{T}}\int_0^\mathbb{T} e^{i \Omega \tau(p-q)} e^{i \mathbf{A}(\tau) \cdot \mathbf{d}_{nm}} d\tau
\label{effhop}
\end{equation}
where $\mathbf{d}_{nm}$ is the vector joining the nearest-neighbor sites in the SPG. $t_{nm}$ is the NNH strength in the absence of light and is assumed to be isotropic that is $t_{nm}=t$. $p$ and $q$ correspond to the band index of Floquet bands. We assume the driving field to be uniform with frequency $\Omega$ and time-period $\mathbb{T}$. Here the vector potential is expressed in units of $ea/c\hslash$ ($a$ being the lattice constant, is taken to be $1 A^o$). 

Finally, with the modified hopping integral, the SPG Hamiltonian (Eq.~\ref{ham-nolight}(a)) can be written as
\begin{eqnarray}
  H_{\text{SPG}} = \sum_n \epsilon_n c_n^\dagger c_n  +\sum_{pq}\left[\left(\sum_{\langle nm \rangle} t_{nm}^{pq} c_n^\dagger c_m + \text{h.c.}\right) - p\hslash\omega\delta_{pq}\right] \nonumber.
\end{eqnarray}
The last term $\left(-p\hslash\omega\delta_{pq}\right)$ originates due to the Fourier transformation of $-i\hslash\partial_t$. The mathematical steps are not shown here to save space. (For a detailed derivation of the modified NNH term and the Hamiltonian, see Refs. ~\cite{gomez-prl,lght1})

\subsection{Two-terminal transmission probability }
We employ the Green's function formalism to calculate the transmission probability of an electron from source to drain through the SPG network. Here, we neglect the Coulomb interaction term and also restrict ourselves within the regime of coherent transport. The effective Green's function can be written as
\begin{equation}
\mathcal{G}^r = \left(E- H_{\text{SPG}}- \Sigma_S -\Sigma_D\right)^{-1}
\end{equation}
where $\Sigma_S$ and $ \Sigma_D$ represent the self-energies of the source and drain, respectively. So, the two-terminal electronic transmission probability can be written in terms of retarded ($\mathcal{G}^r$) and advanced $\left(\mathcal{G}^a\left(=\mathcal{G}^r\right)^\dagger\right)$ Green's functions as
\begin{equation}
\mathcal{T}= \text{Tr}\left[\Gamma_S \mathcal{G}^r \Gamma_D \mathcal{G}^a \right]
\end{equation}
where $\Gamma_S$ and $\Gamma_D$ are the coupling matrices that describe the rate at which particles scatter between the leads and the fractal network.

\subsection{Thermoelectric quantities}
Thermoelectric materials convert heat into electric energy and vice versa. The heat-to-electric energy conversion efficiency is
expressed by a dimensionless quantity, known as the {\it figure of merit} (FOM) which is denoted by $ZT$. The expression of FOM is given by
\begin{equation}
ZT = \frac{G S^2 T}{k(=k_e+k_{ph})}
\label{Eq-fom}
\end{equation}
where $G$ is the electronic conductance, $S$ is the Seebeck coefficient (thermo power), and $T$ is the temperature. $k$ represents the total thermal conductance which is a sum of electronic conductance $(k_e)$ and phononic conductance $(k_{ph})$. Each of these quantities, apart from $k_{ph}$, can be calculated from Landauer prescription~\cite{dubi,lam} as
\begin{subequations}
\begin{eqnarray}
G & =&\frac{2 e^2}{h}L_0 \\
S & =&- \frac{1}{eT} \frac{L_1}{L_0}\\
k_e & = & \frac{2}{hT} \left(L_2 - \frac{L_1^2}{L_0}\right).
\end{eqnarray}
\label{gsk}
\end{subequations}
In the above expressions, the Landauer integrals $L_n$ are defined as
\begin{equation}
L_n= - \int \mathcal{T}(E) (E-E_f)^n \frac{\partial f_{FD}}{\partial E} dE
\label{ln}
\end{equation}
where $E_f$ is the fermi energy, $\mathcal{T}(E)$ is the transmission probability of the system, and $f_{FD}(E)$ represents the Fermi-Dirac distribution function. Typically $ZT > 1$ is regarded to be a good thermoeletric material. However, for large-scale energy-conversion systems $ZT \sim 2-3$  is often prescribed~\cite{Tritt-Annu-Rev-Mat-Res}. 
\subsection{Phonon thermal conductance} The phononic contribution to the thermal conductance $ k_{ph}$ is often neglected as an approximation in Eq.~\ref{Eq-fom}. This is due to the fact that at nanoscale regime, the system contains less number of lattice sites, and therefore, at low or even moderate temperatures, the contribution is relatively small. But for precise estimation of $ZT$, one needs to include $ k_{ph}$. When the temperature difference between the two contact electrodes is infinitesimally small, the phonon thermal conductance is evaluated from the expression~\cite{zhang-ph,hopkins-prb,aghosh}
\begin{equation}
k_{ph}= \frac{\hslash}{2\pi}\int_0^{\omega_c} \mathcal{T}_{ph}\frac{\partial f_{BE}}{\partial T}\omega d\omega .
\end{equation}
Here, $\omega$ is the phonon frequency $\omega_c$ is the cut-off frequency. Here we consider only elastic scattering. $f_{BE}$ is the Bose-Einstein distribution function. $\mathcal{T}_{ph}$ is the phonon transmission coefficient across the SPG, and it is computed using the well known Green's function prescription through the relation
\begin{equation}
\mathcal{T}_{ph}= \text{Tr}\left[\Gamma_S^{ph} \mathcal{G}_{ph} \Gamma_D^{ph} \left(\mathcal{G}_{ph}\right)^\dagger \right]
\end{equation}
$\Gamma_{S/D}^{ph}=i\left[\widetilde{\Sigma}_{S/D}-\widetilde{\Sigma}_{S/D}^\dagger\right]$ is the thermal broadening and $\widetilde{\Sigma}_{S/D}$ is the self-energy matrix for the source/drain electrode. The Green's function for the SPG can be written as
\begin{equation}
G_{ph} = \left[{\mathbb M}\omega^2 - {\mathbb K} -\widetilde{\Sigma}_S - \widetilde{\Sigma}_D\right]
\end{equation}
where ${\mathbb M}$ is a diagonal matrix representing the mass matrix of the SPG. A diagonal element ${\mathbb M}_{nn}$ of this matrix denotes the mass of the atom at the $n$-th position in the SPG. ${\mathbb K}$ is the matrix of spring constants in the SPG. The element ${\mathbb K}_{nn}$ represents the restoring force of the $n$-th atom due to its neighboring atoms, whereas the element ${\mathbb K}_{nm}$ describes the effective spring constant between $n$-th atom and its $m$-th neighboring atom. The self-energy matrices $\widetilde{\Sigma}_S$ and $\widetilde{\Sigma}_D$ have the same dimension as ${\mathbb M}$ and ${\mathbb K}$ and can be computed by evaluating the self-energy term $\Sigma_{S/D}=-K_{S/D}\,\text{exp}\left[2i\,\text{sin}^{-1}\left(\frac{\omega}{\omega_c}\right)\right]$, where $K_{S/D}$ is the spring constant at the electrode-SPG contact interface.

The spring constants are calculated from the second derivative of Harrison's interatomic potential~\cite{harrison}. For the 1D electrodes, the spring constant is given by $K=3dc_{11}/16$, while for the SPG $K=3d\left(c_{11}+2c_{12}\right)/16$. Here $d$ denotes the interatomic spacing and $c_{11}$ and $c_{12}$ are the elastic constants. The difference in the expressions of the spring constants arises because, in a 1D electrode, there is no transverse interaction, but in a 2D system like SPG, one needs to consider it~\cite{kittel}. With the knowledge of the mass and spring constant, the cut-off frequency for the 1D electrode is determined from the relation $\omega_c=2\sqrt{K/M}$. For a detailed description of the procedure to calculate the phonon thermal conductance, see Refs.~\cite{zhang-ph,hopkins-prb,aghosh}.


\section{Results}

Before discussing the results, let us first mention the parameters used in the present work. The on-site energies in the fractal network as well as in the source and drain electrodes are set at zero. All the energies reported here are measured in units of electron-volt (eV). The NNH hopping integrals are considered in the wide-band limit, where it is set for the electrodes as $t_0=2\,$eV, and in the SPG as $t=1\,$ eV. The coupling strengths of the SPG to the source and drain electrodes, characterized by the parameters $\tau_S$ and $\tau_D$, are also fixed at $1\,$eV. For any other choice of parameter values, the physical picture will be qualitatively the same, which we confirm through our exhaustive calculation. The rest of the parameter values, which are not common for the entire analysis, are mentioned in the appropriate places of our analysis.

Due to the time-periodicity of the driving field, a periodically driven ${\mathbb D}$-dimensional lattice is equivalent to an undriven ${\mathbb D}+1$-dimensional lattice~\cite{gomez-prl,lght1}. For such a periodically driven system, the initial Bloch band breaks into Floquet-Bloch (FB) bands, where the coupling between FB bands depends directly on the driving frequency regime. This time-independent ${\mathbb D}+1$ dimensional lattice can be visualized as if the SPG is connected to its several virtual copies arranged vertically to the lattice plane. In the high-frequency limit, the Floquet bands decoupled from each other, and only the zeroth-order Floquet band ($p=q=0$) has the dominant contribution in Eq.~\ref{effhop}, while other higher-order terms in $p$ and $q$ essentially have a vanishingly small contribution. Due to this decoupling process, the coupling between the parent SPG lattice and its virtual copies becomes vanishingly small. This scenario is no longer valid in the low-frequency regime, where the virtual copies are directly coupled to the parent SPG lattice. Therefore, in the low-frequency limit, several virtual copies of the SPG lattice come into the picture. Consequently, the effective size of the system increases. This could decrease the phase-relaxation length, and at finite temperature, it might be reduced further. Thus, it will be quite hard to get favorable transport properties in the low-frequency regime.

Because of the above facts, we restrict the present analysis in the high-frequency limit $\hslash \omega \gg 4t$. The light frequency for this limiting case should be at least $\sim 10^{15}\,$Hz, which is in the near-ultraviolet/extreme ultraviolet regime. The corresponding electric field $\sim 10^4\,$V/m, while the magnetic field is $\sim 10^{-5}\,$T. Since the magnetic field due to the light irradiation is vanishingly small, its effect can safely be ignored. The intensity of the light irradiation is $\sim 10^5\,$W/m$^2$, and is certainly within the experimental reach. Since much higher light intensities have been used in several other recent works~\cite{cwd1,cwd2}, we strongly believe that our chosen intensity will no longer damage the physical system.

\subsection{Detection of mobility edge}
\begin{figure}[ht!]
\centering
\includegraphics[width=0.35\textwidth]{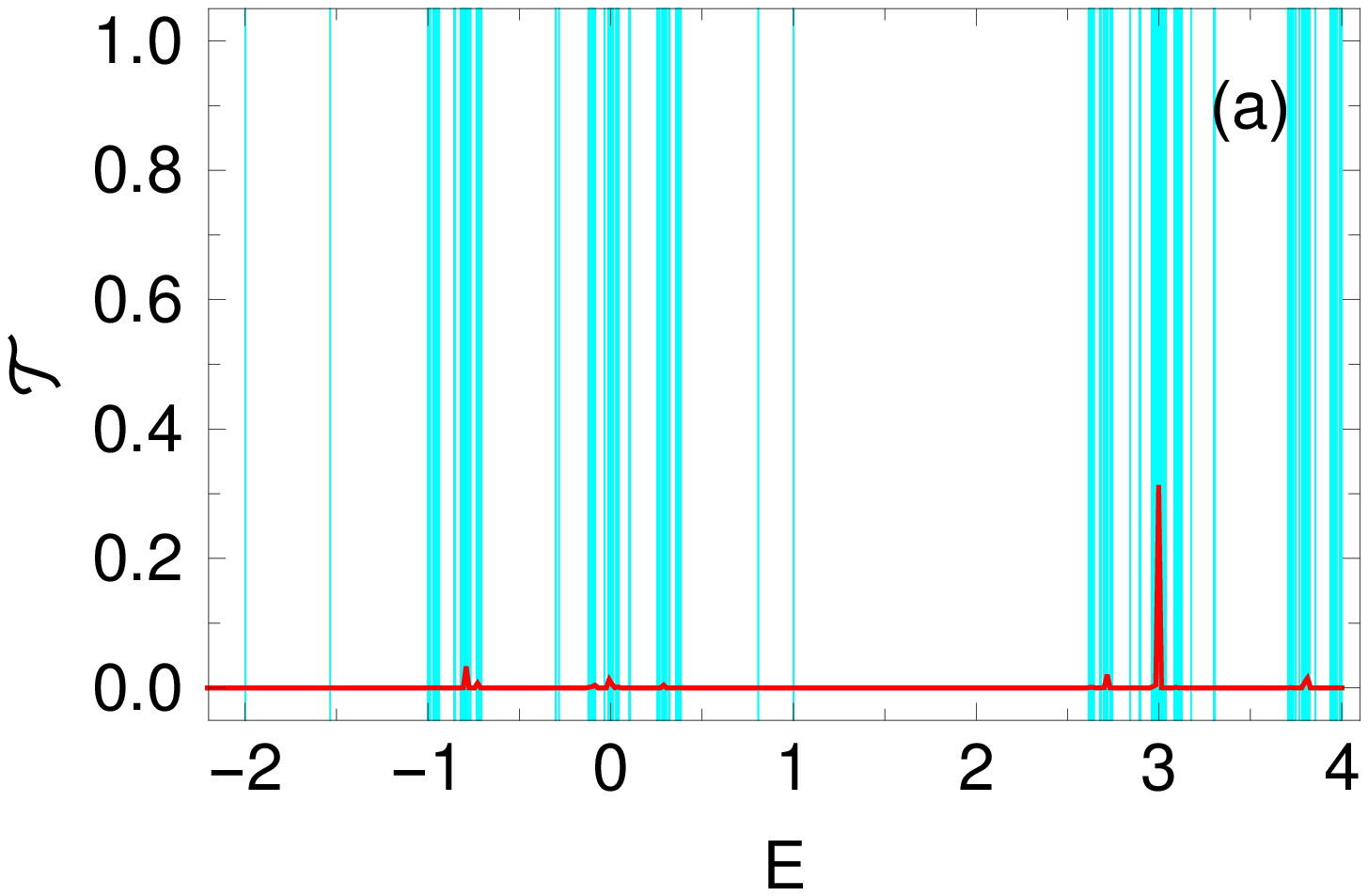}\vspace{0.5cm}
\includegraphics[width=0.35\textwidth]{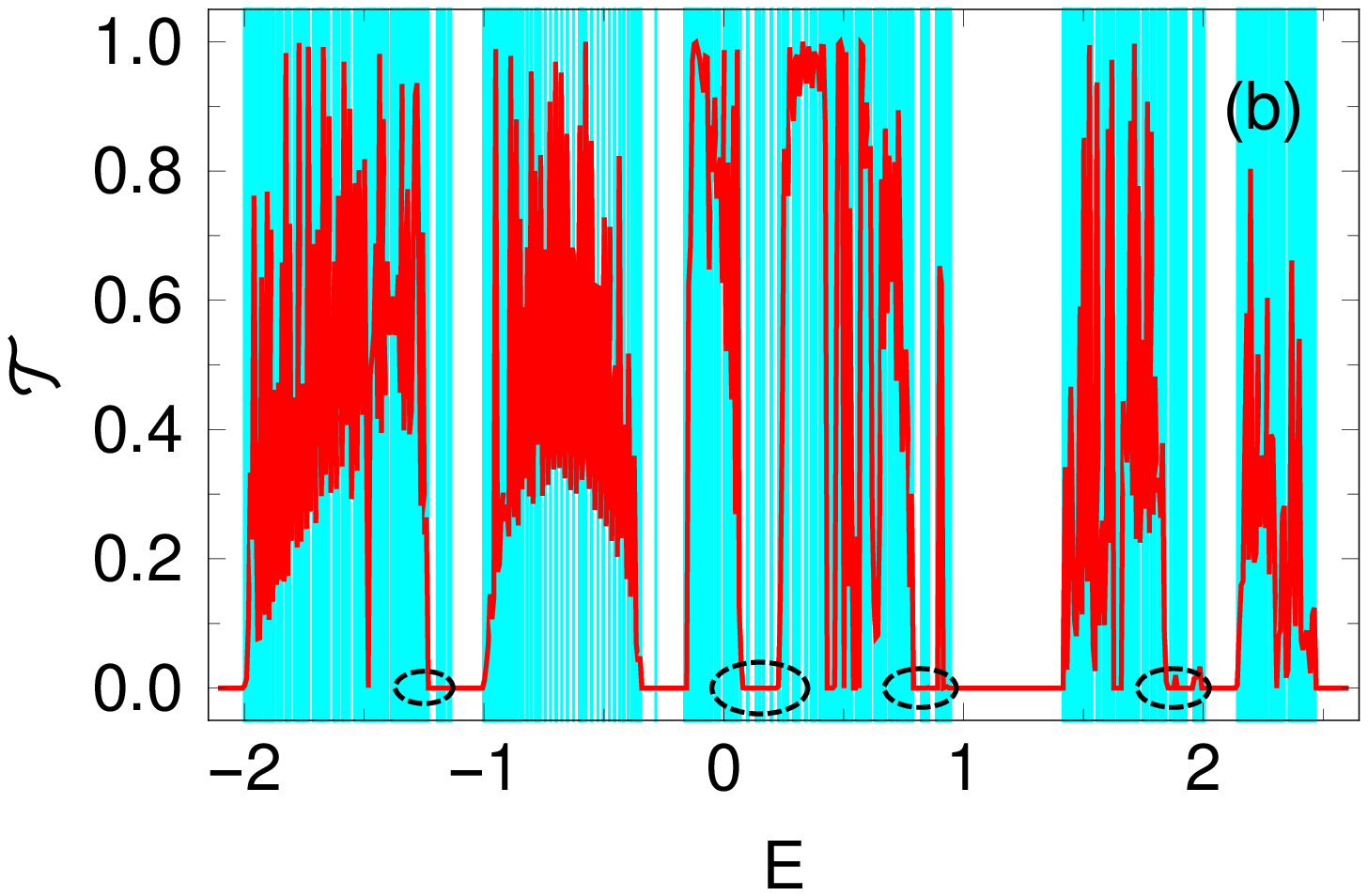}
\caption{(Color online).  Transmission probability $\mathcal{T}(E)$ (red color) as a function of energy along with the energy eigenvalues (cyan color) for an 8th generation SPG network, where (a) and (b) correspond to the results in the absence and presence of light respectively.  At each eigenvalue, we draw a single vertical line of unit magnitude for better visibility of the energy eigen spectrum. The light parameters are $A_x =0$ and  $A_y = 2$.}
\label{gen8}
\end{figure}

We begin our discussion by analyzing the two-terminal transmission coefficient along with the energy eigenvalues of an SPG network in the absence and presence of light, as shown in Fig.~\ref{gen8}. The transmission spectrum (red color) is superimposed on the spectrum of energy eigenvalues, where we draw a vertical line of unit magnitude (cyan color) in each of these eigenvalues. To check the localization behavior in the asymptotic limit, we consider a bigger SPG (8th generation) that contains a fairly large number of lattice sites. Figure~\ref{gen8}(a) shows that the eigenenergies are highly degenerate. Few sub-bands are formed, providing finite gaps. Along with the bands, some isolated energy eigenvalues are also visible.
Here, the transmission coefficient becomes zero or vanishingly small for the entire allowed energy window, which emphasizes a complete localization of all the states. This localized behavior for the irradiation-free SPG is known in the literature. 

The situation becomes quite interesting 
and important as well when the system is irradiated. Figure~\ref{gen8}(b) shows that degeneracies get removed, and thus wider bands are formed. The most striking feature is observed in the transmission spectrum, where almost all the energy levels are associated with finite transmission probabilities. Thus, it gives a clear indication of a localization to delocalization transition in the presence of light irradiation. Now, a careful inspection reveals that near the energy $E \sim 0$, there is a fine strip of eigenvalues that are completely localized since the transmission coefficient is identically zero within this fine strip. To the immediate left/right of the strip, the transmission coefficient is finite which clearly manifests that the states are extended. 
Thus, we have a sharp edge that separates the extended and localized energy eigenstates which validates the existence of a mobility edge in the presence of light in SPG. Interestingly, we find that such a mobility edge appears at multiple energies like $E\sim-1,1$, and 2, etc. The region across a mobility edge is marked by a black dotted ellipse in Fig.~\ref{gen8}(b) for better visualization.

Another interesting feature we observe here is that in the presence of light, the allowed energy window gets reduced ($ \sim -2 $ - $ 2.5$) than in the absence of light ($\sim -2 $ - $ 4$). We shall talk about this feature at length in the next subsection. Overall, what we accumulate is that localization to delocalization transition is obtained in the presence of light along with multiple mobility edges.

\subsection{Spectral analysis}

\begin{figure}[ht!]
\includegraphics[width=0.23\textwidth]{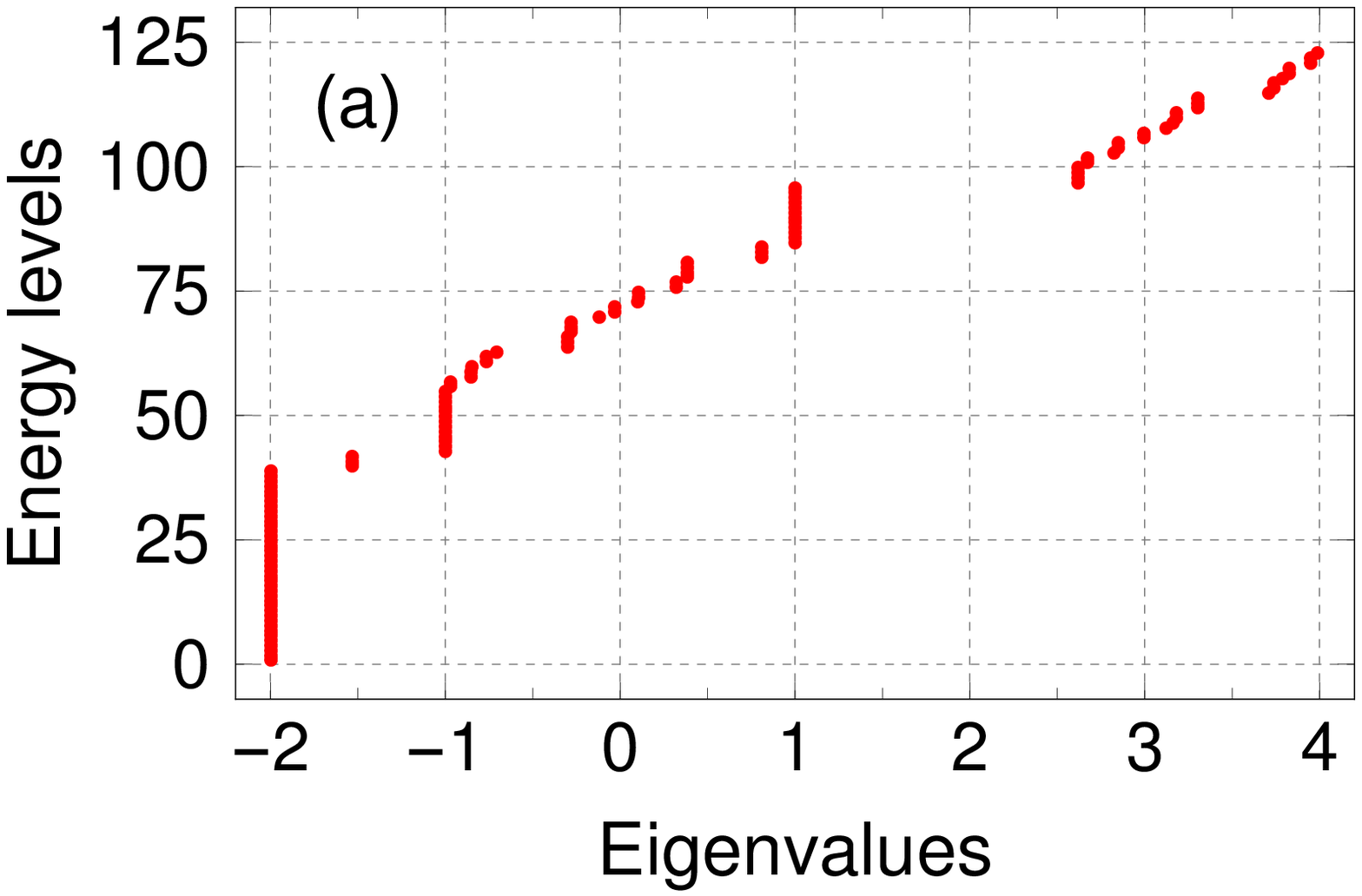} \hfill
\includegraphics[width=0.23\textwidth]{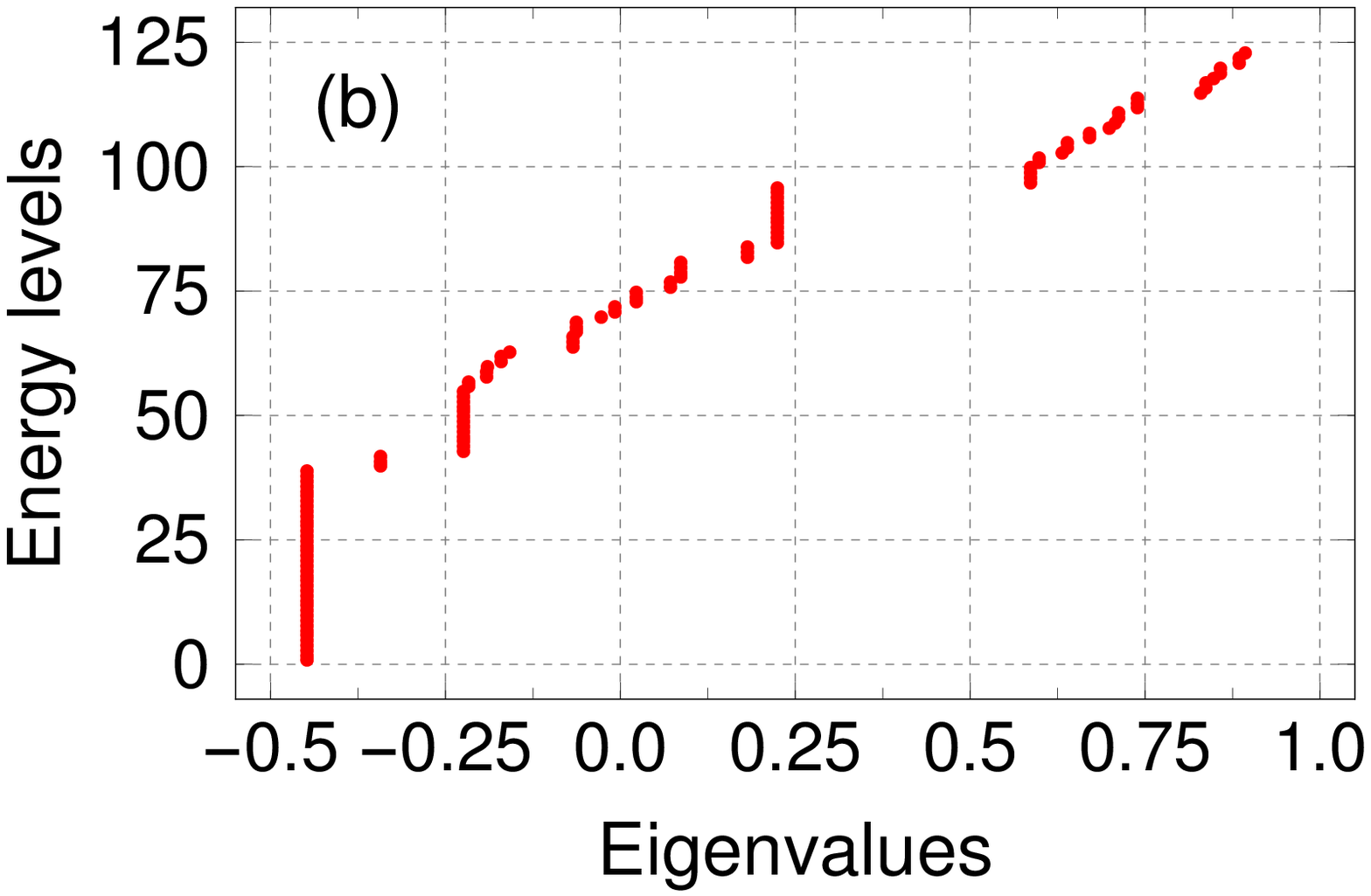}\vskip 0.1 in
\includegraphics[width=0.23\textwidth]{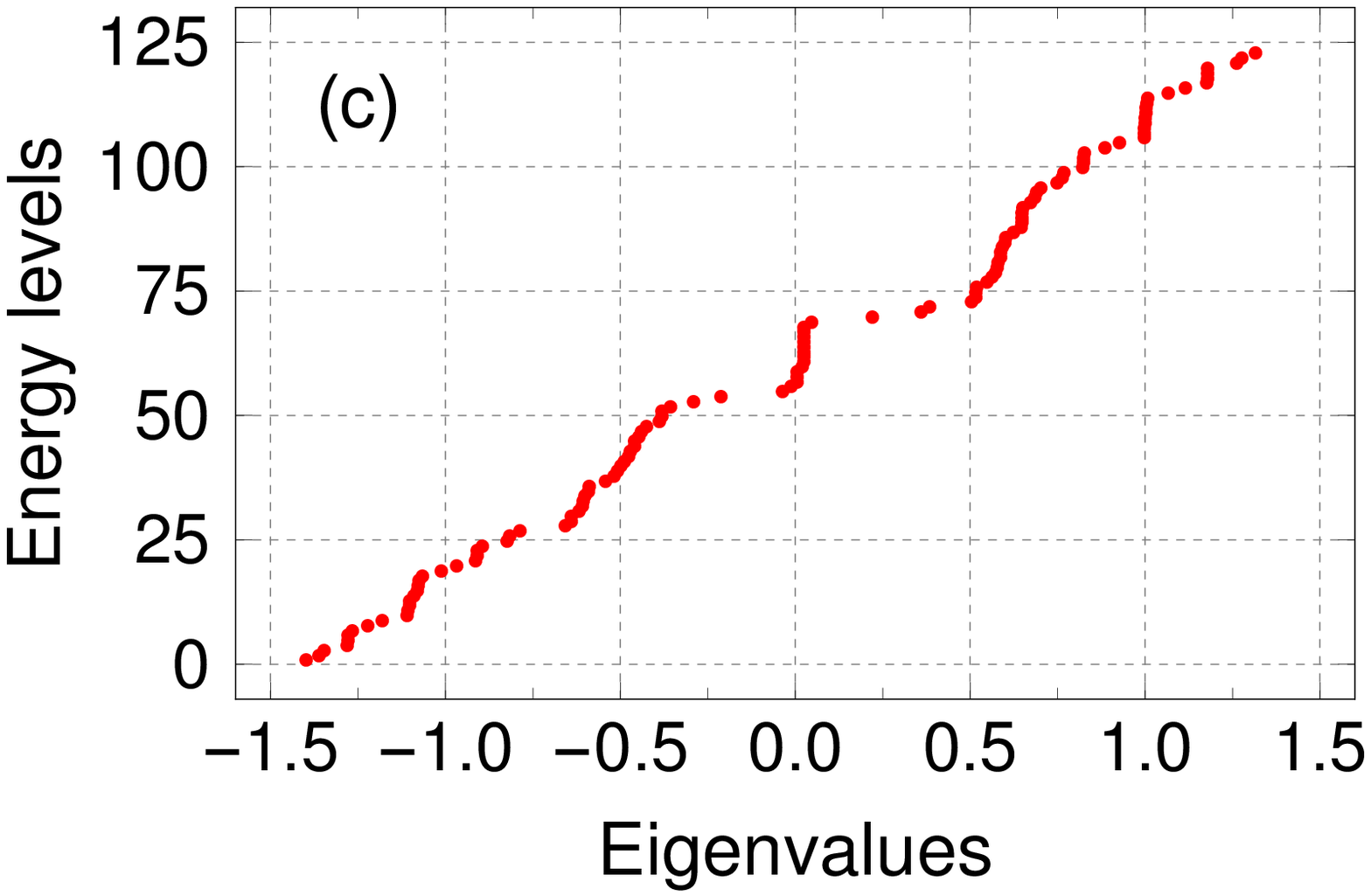} \hfill
\includegraphics[width=0.23\textwidth]{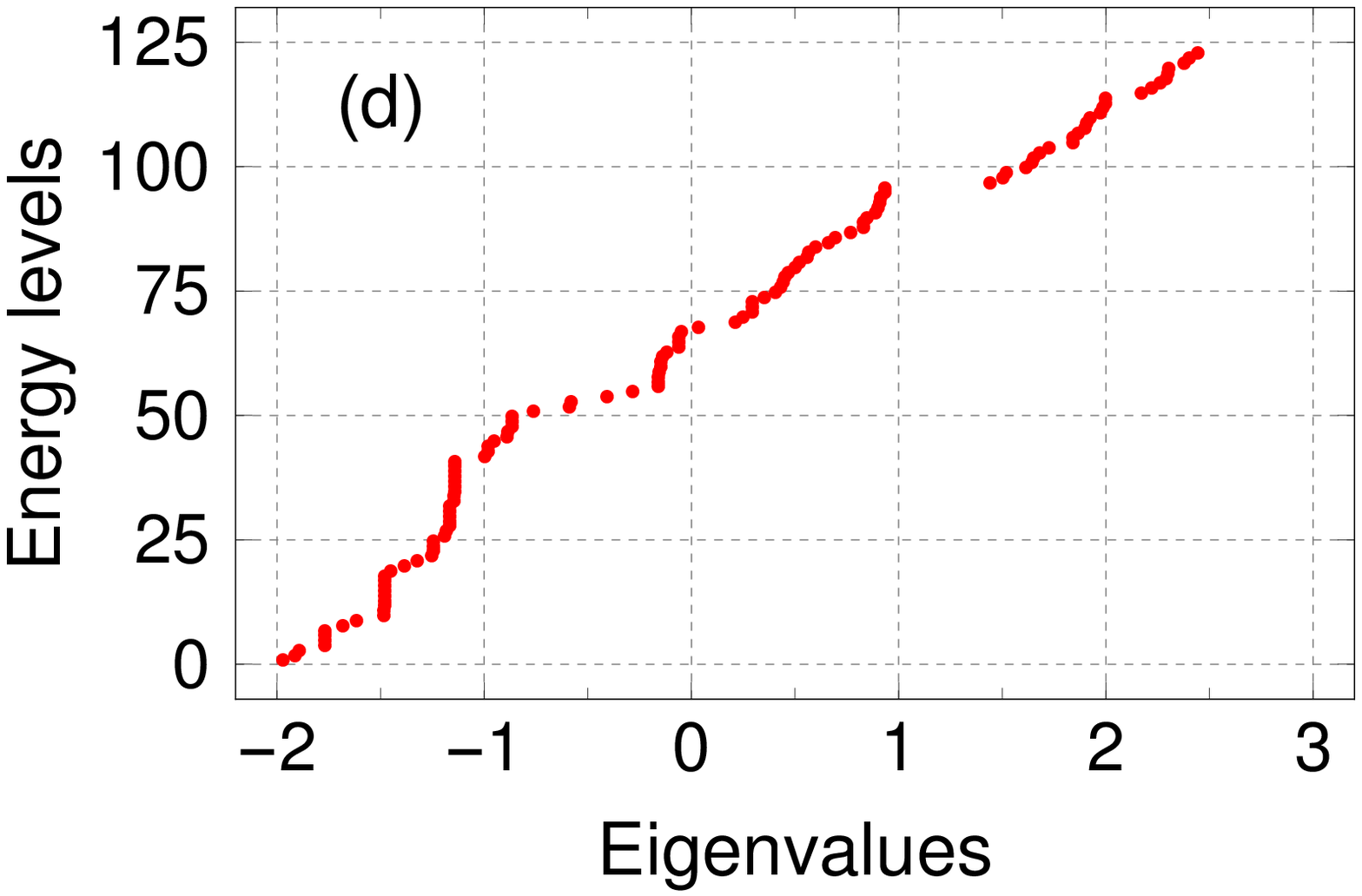}
\caption{(Color online). Eigenvalue spectra  in the absence and presence of light with different irradiation parameters. (a) Absence of light, (b) $A_x = A_y = 2$ and $\phi = \pi/2$, (c) $A_x = A_y = 2$ and $\phi = \pi/4$, and (d) $A_x = 0$ and $A_y = 2$.}
\label{ELvsEV}
\end{figure}

Now, we discuss the spectral behavior of an SPG network, which are extremely crucial to understand the electronic transport phenomena. Here, we consider a 5th generation SPG to analyze the results. In Fig.~\ref{ELvsEV}, we show the spectra of energy eigenvalues both in the presence and absence of irradiation. In the absence of light, the spectrum is highly degenerate and gapped, as shown in Fig.~\ref{ELvsEV}(a). 
These are the basic characteristics of any fractal geometries and are well-known in literature.~\cite{domany,maiti-pla}. In the case of a circularly polarized (CP) light, the nature of the spectrum is identically the same like what we find in the irradiation-free case, but the allowed energy window gets shortened as depicted in Fig.~\ref{ELvsEV}(b). The identical energy spectrum and reduced energy window for the case of CP light can be explained as follows. For the SPG network, there are two hopping directions, one is the horizontal hopping, and the other is the angular one. In the presence of an arbitrarily polarized light, the hopping strengths in both directions are modified according to Eq.~\ref{effhop} by the Bessel function of the first kind~\cite{gomez-prl,sudin-jpcm-cc}, but their strengths become the same for the CP light.

 Therefore, the hopping strength is isotropic for the CP light, like the case in the absence of light but with a reduced hopping terms. This isotropic nature of the hopping integrals makes the spectrum identical with that of the irradiation-free case, and the reduced hopping term is responsible for the decrease in the allowed energy window. For an elliptically polarized light (the horizontal and angular hopping strengths are different now), the degeneracy is broken, and the energy levels are regularly arranged (Fig.~\ref{ELvsEV}(c)). Here the allowed energy window is also different than the previous two cases. For the linearly polarized light as shown in Fig.~\ref{ELvsEV}(d), the broken degeneracy levels with modified allowed energy windows are observed due to the modified NNH integrals in both the directions.

\begin{figure}[ht!]
\includegraphics[width=0.23\textwidth,height=0.15\textwidth]{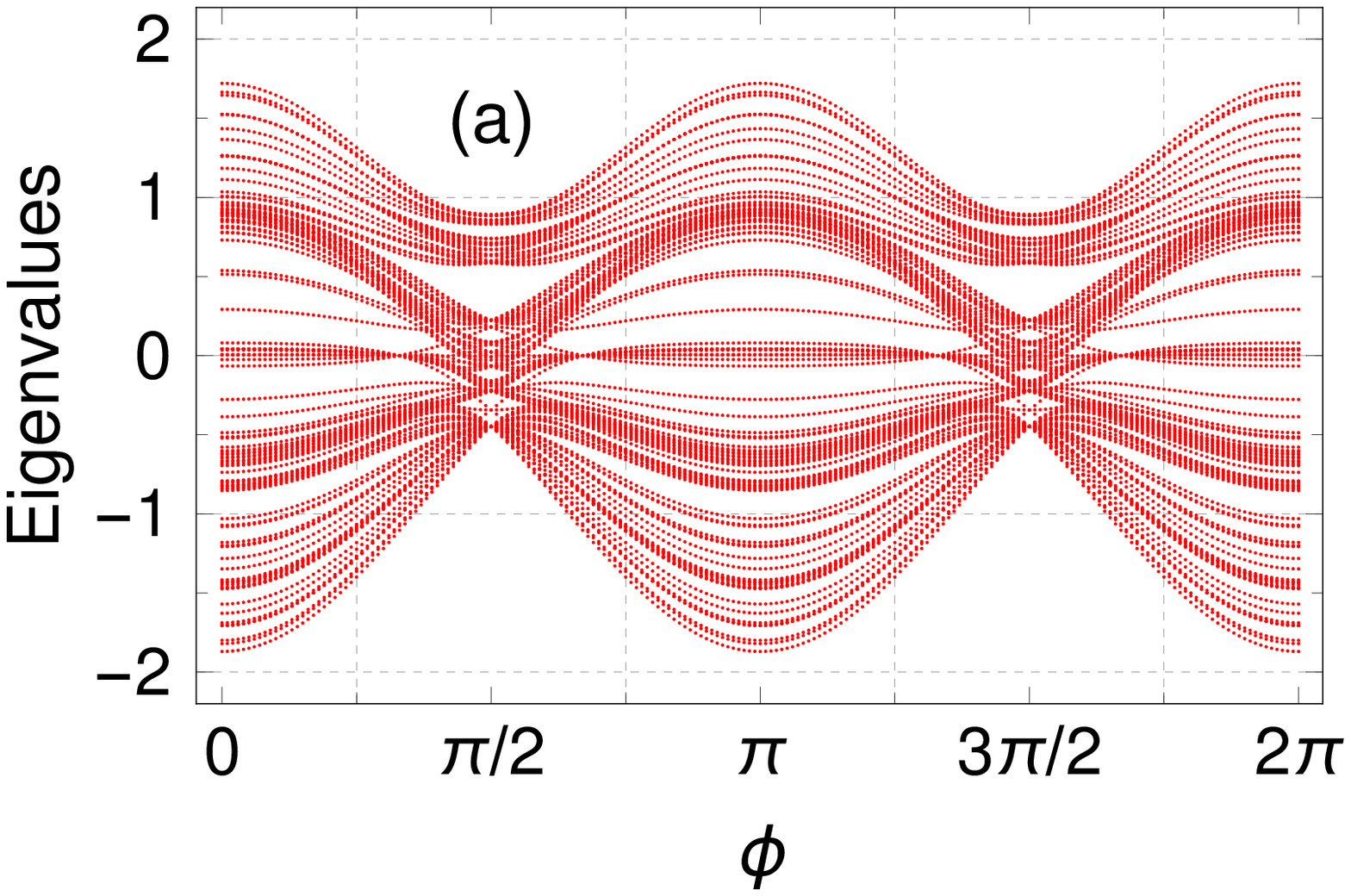} \hfill
\includegraphics[width=0.23\textwidth,height=0.15\textwidth]{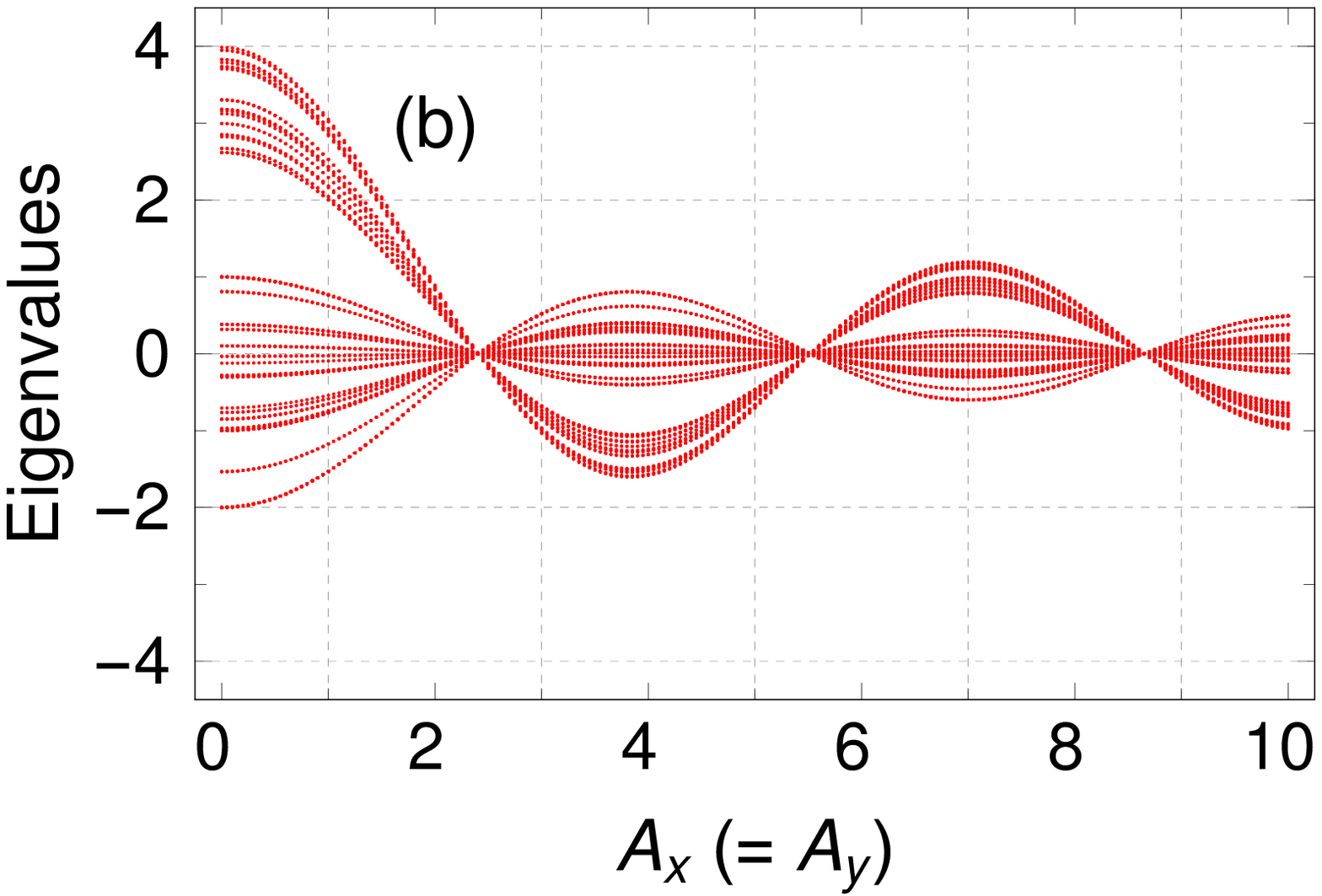}
\caption{(Color online). Eigenvalues as a function of (a) phase $\phi$ with $A_x = A_y = 2$ and (b) field amplitude $A_x (=A_y)$ with $\phi = \pi/2$.}
\label{EVvsLP}
\end{figure}

To have a better understanding of the effect of light irradiation, in Fig.~\ref{EVvsLP} we examine the spectrum of energy eigenvalues in terms of the light parameters. Figure~\ref{EVvsLP}(a), shows the eigenvalues as a function of phase of the vector potential $\phi$ for $A_x=A_y =2$. The allowed energy window oscillates 
with a period $\phi=\pi/2$. Therefore, it is possible to control the energy window externally by adjusting the light parameter $\phi$, which is useful in engineering transport phenomena that can be understood from our forthcoming analysis.

Figure~\ref{EVvsLP}(b) describes the behavior of energy eigenvalues with field amplitudes for CP light that is by varying $A_x \left(=A_y\right)$ keeping phase fixed at $\phi = \frac{\pi}{2}$. The observation is quite remarkable in the sense that the allowed energy window can significantly be modified by tuning the field amplitude. For instance, in the absence of light, the eigenvalues are distributed within the interval -2 to 4. Once we irradiate the SPG with CP light, the allowed window gets shortened with increasing the field amplitude, and around $A_x=A_y=2.4$, all the eigenvalues coincide exactly at zero energy. This particular feature could be useful in switching applications due to the following reason. Suppose we are measuring the electrical conductance, setting the Fermi energy apart from zero. In the absence of light, the conductance is finite. Now, in the presence of CP light, when the field amplitude is around 2.4, there will be no such state at that energy, and consequently, the conductance will be zero. Beyond $A_x=A_y=2.4$, we note another two values of the field amplitude where all the eigenvalues again coincide to zero. A band inversion is also observed for $A_x>2.4$. This is due to the modified hopping term, which stays positive up to $A_x>2$, and beyond that, it becomes negative. As the phase of the hopping integral reveres, the band inversion takes place. Earlier in Fig.~\ref{ELvsEV}(b), we claimed that in the presence of CP light, the fractal nature of the energy spectrum remains intact with a modified bandwidth. This is true for any other field amplitude, including the band inversion cases where only the structure of the energy spectrum gets inverted. This claim also holds for the field amplitudes where all the eigenvalues coincide at zero energy which we confirm through our exhaustive analysis. Here we would like to note that the energy eigenvalues coincide to zero as we set $\epsilon_n=0$. For the situation, when $\epsilon_n$ is finite, all the eigenvalues will coincide to that particular value without altering the characteristic features.

\begin{figure}[ht!]
\begin{center}
\includegraphics[width=0.35\textwidth]{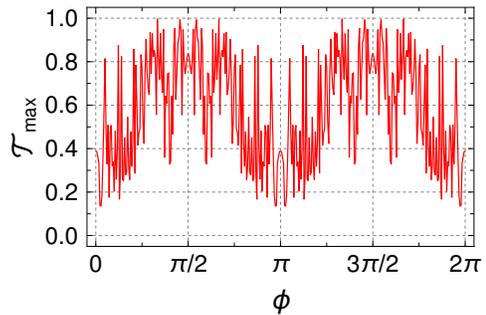}
\caption{(Color online). The maximum transmission coefficient $\mathcal{T}_{max}$ as a function of $\phi$ with $A_x=A_y=2$.}
\label{tvsphi}
\end{center}
\end{figure}

In Figs.~\ref{ELvsEV}, and \ref{EVvsLP}, we find that the eigenenergies are significantly modified in the presence of light. Thus, it will be interesting to study how the transmission coefficient $\mathcal{T}$ behaves with the light parameters since $\mathcal{T}$ plays a crucial role in transport phenomena. In Fig.~\ref{tvsphi},  
we plot the maximum of the transmission coefficient, $\mathcal{T}_{max}$ as a function of $\phi$. $\mathcal{T}_{max}$ is evaluated by taking the maximum of $\mathcal{T}$ within the allowed energy window for a given light parameter. Here $\mathcal{T}_{max}$ shows oscillatory behavior similar to the eigenvalue spectrum with a period $\phi=\pi/2$. Interestingly, $\mathcal{T}_{max}$ varies in a wide range (0.2 to unity), and therefore,  the conducting behavior of the SPG network can be adjusted by regulating the phase factor.

So far, several interesting features have emerged from the study of the energy spectrum and transmission characteristics, such as the localization to delocalization phenomena, the existence of multiple mobility edges, breaking of the level degeneracy, the possibility to engineer the bandwidth, etc. Now, we try to exploit the delocalization phenomena and the existence of the multiple mobility edges (Fig.~\ref{gen8}) in TE application for this quantum network.

\subsection{Thermoelectric properties}
\begin{figure}[ht!]
\includegraphics[width=0.23\textwidth,height=0.15\textwidth]{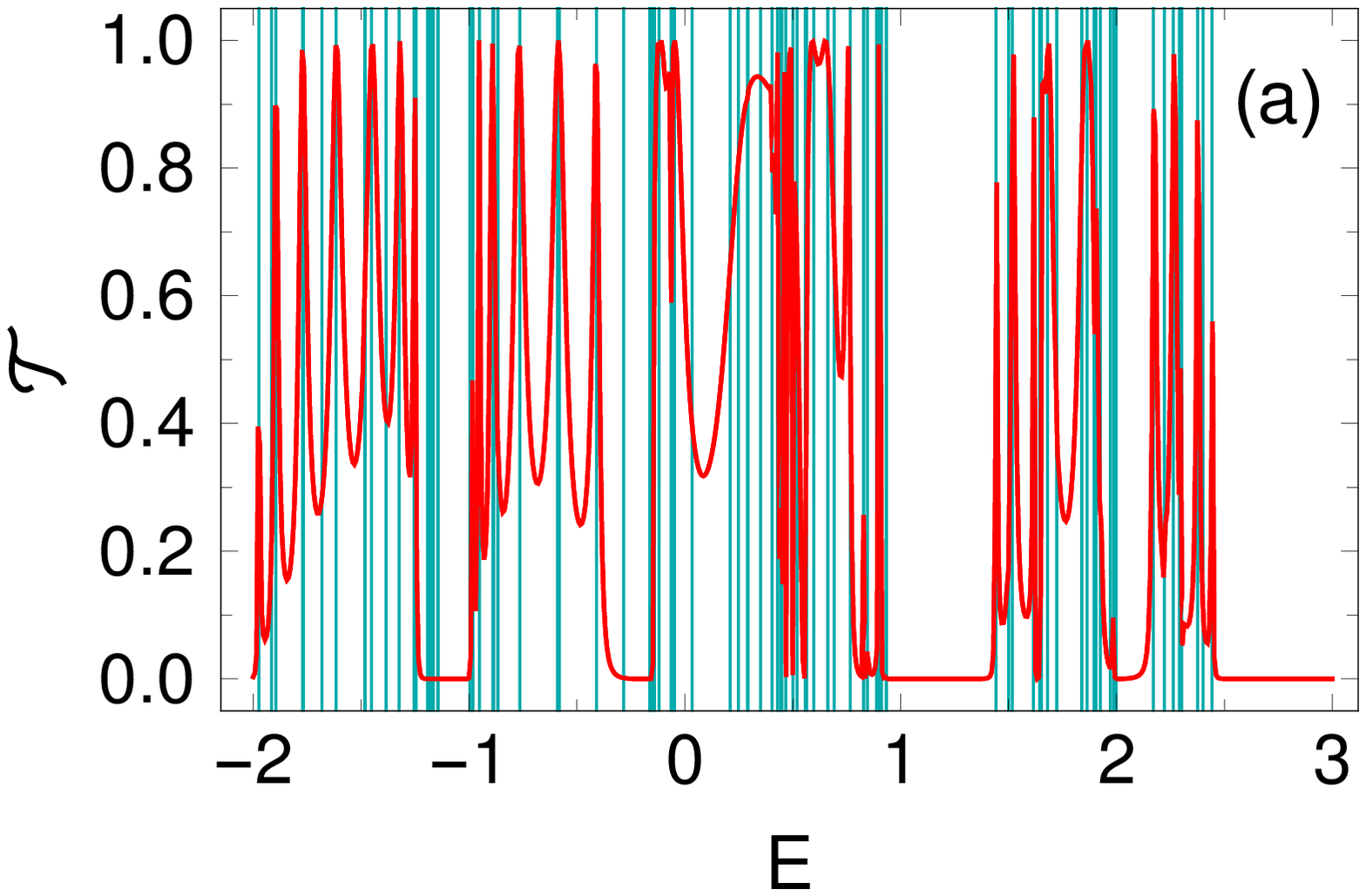} \hfill
\includegraphics[width=0.23\textwidth,height=0.15\textwidth]{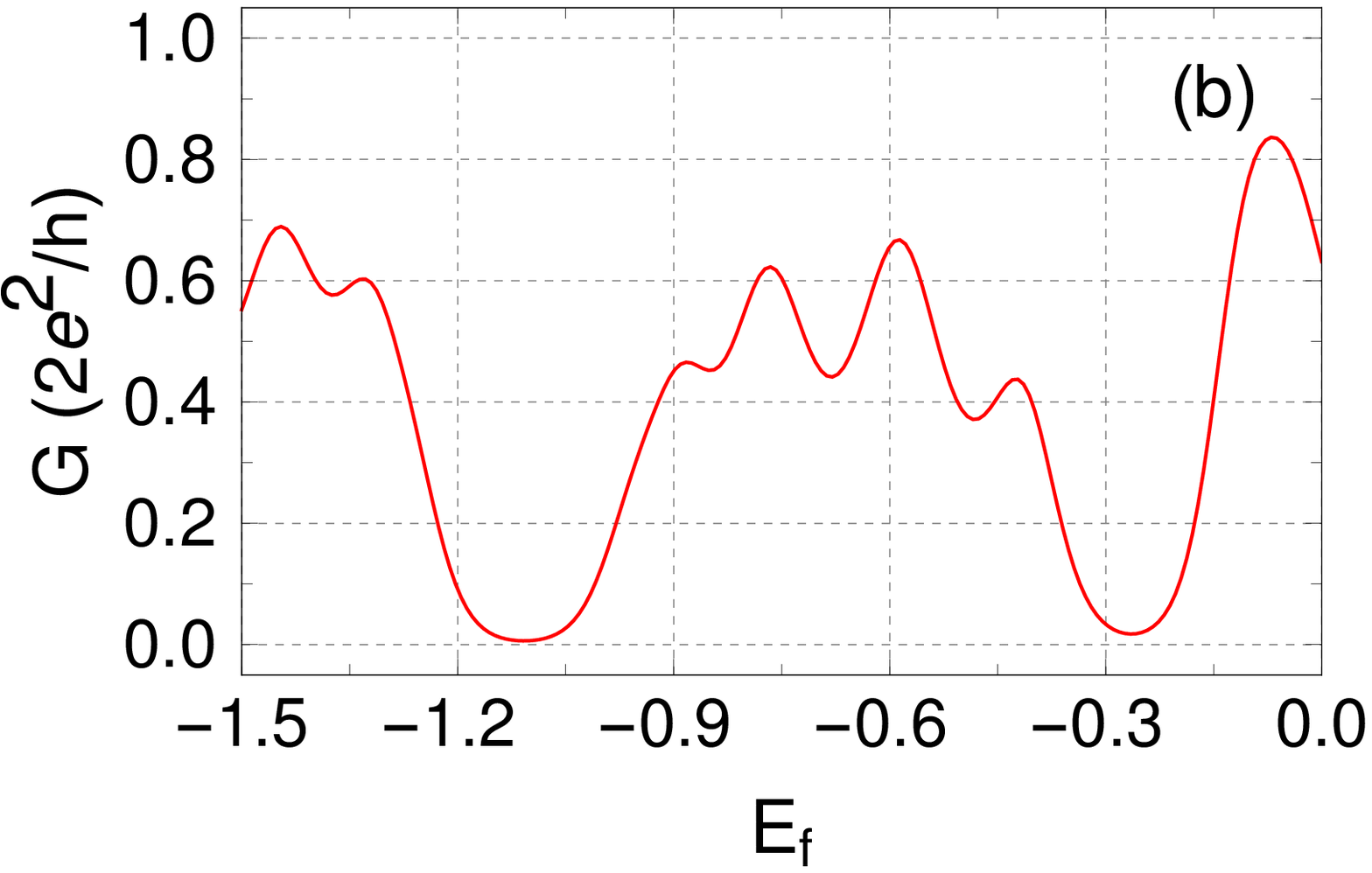}\vskip 0.1 in
\includegraphics[width=0.23\textwidth,height=0.15\textwidth]{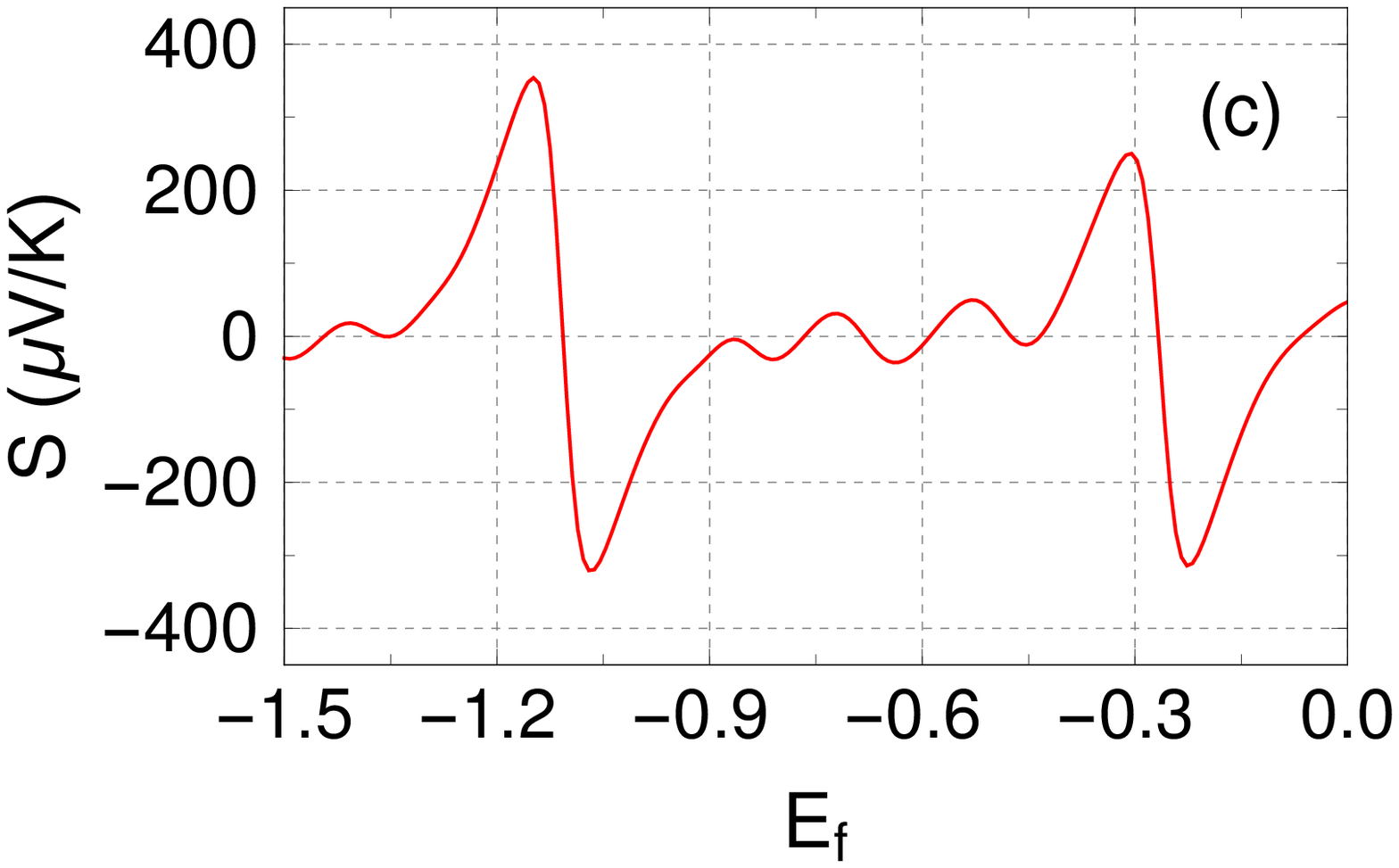}\hfill
\includegraphics[width=0.23\textwidth,height=0.15\textwidth]{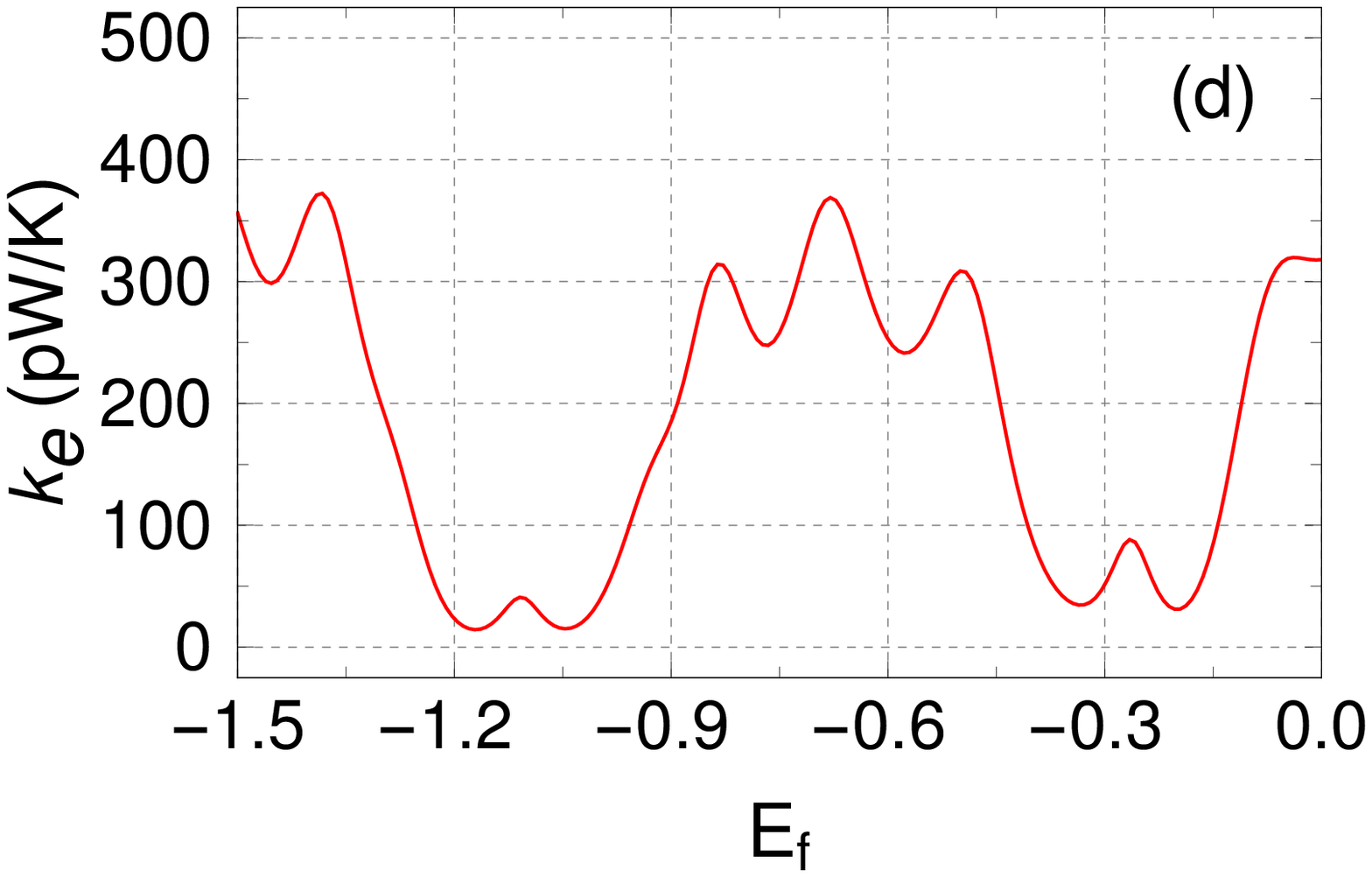} 
\caption{(Color online).  (a) Transmission coefficient $\mathcal{T}$ (red color) as a function of energy $E$ along with the energy eigenvalues (cyan color). At each eigenvalue, we draw a single vertical line of unit magnitude for better visibility of the energy eigen spectrum. (b) Electrical conductance $G$, (c) thermopower $S$, and (d) thermal conductivity due to electrons $k_{e}$ as a function of the Fermi energy $E_F$. Here a 5th generation SPG is considered and the light parameters are $A_x=0$ and  $A_y=2$.}
\label{tgsk}
\end{figure}

Unless mentioned otherwise, all the TE quantities are evaluated at room temperature $T=300\,$K. We consider a 5th generation SPG for demonstration and the light parameters are $A_x=0$ and $A_y=2$. In Fig.~\ref{tgsk}(a), the transmission spectrum (red color) is superimposed on the spectrum of energy eigenvalues (cyan color). At each eigenvalues, we draw a vertical line of unit magnitude for better visualization.

The transmission spectrum shows almost similar behavior with energy that was obtained for the 8th generation (Fig.~\ref{gen8}) with the identical light parameters owing to the self-similar structure of SPG. Across $E\sim-1$, a mobility edge is found, similar to what we get for the 8th generation SPG, and we vary the Fermi energy window across this energy. Now, we compute the electrical conductance, thermopower, and thermal conductance due to electron as a function of the Fermi energy as shown in Figs.~\ref{tgsk}(b), (c), and (d) respectively. The $G$-$E_F$ spectrum basically follows the $\mathcal{T}$-$E$ curve as $G$ is evaluated using Eqs.~\ref{gsk}(a) and \ref{ln}. The spikes in the transmission spectrum smeared out due to the temperature broadening. We observe a dip around the mobility edge in Fig.~\ref{tgsk}(a). On the other hand, the thermopower becomes maximum around the mobility edge (Fig.~\ref{tgsk}(b)), acquiring a value $\sim 380\,\mu$V/K. The behavior of the electronic thermal conductance $k_{ph}$ is similar to that of $G$ and is of the order of a few hundreds of pW/K.

The suppression of $G$ and $k_e$ around the mobility edge is simply due to the sharp fall in the transmission spectrum. The asymmetry in the transmission function around the mobility edge is responsible for the enhancement of $S$. The thermopower is evaluated using Eq.~\ref{gsk}(b) and the corresponding thermal integral $L_1$ (Eq.~\ref{ln}), where the transmission spectrum $\mathcal{T}(E)$ is weighted by the terms $(E-E_f)$ and $\left(\partial{f}_{FD}/\partial{E}\right)$. The latter term provides the broadening and is antisymmetric around $E_F$. Thus, if $\mathcal{T}(E)$ is symmetric around $E_F$, then $S$ will be zero irrespective of the value of the transmission. Therefore, to get higher thermopower, we need an asymmetric transmission function that can be easily obtained across the mobility edge.

So, what we gather from Fig.~\ref{tgsk} is that both $G$ and $k_{ph}$ decrease, and $S$ increases around the mobility edge. Though $G$ is directly proportional to $ZT$, the decrease in $G$ does not suppress the FOM considerably since $k_{e}$ comes in the denominator in the expression of FOM (Eq.~\ref{Eq-fom}). As $ZT\propto S^2$, a small increase in the thermopower enhances the FOM significantly. Therefore, all the TE results indicate a high $ZT$ around the mobility edge. However, before we make any definite conclusion regarding the efficiency,  it is important to study the thermal conductance due to phonon, which we discuss now.

\noindent{\bf Phonon thermal conductance:}  Before discussing the results due to phonons let us briefly mention the values those are considered for our calculation. We describe the system to calculate $k_{ph}$ as a spring-mass system. The spring constants are usually calculated from the second derivation of the harmonic Harrison potential~\cite{harrison}.    For the 1D electrodes, the spring constant is considered as $16.87\,$N/m, while in the SPG, the spring constant is $23.93\,$N/m. The considered values of the sping constants are corresponding to typical semiconductors such as Ge and Si~\cite{aghosh}. The cut-off frequency of vibration depends on the material properties of the electrodes and the SPG since we assume that two different atoms are adjacent to each other in the present work. In our chosen setup, by averaging the spring constants of the electrodes and SPG, and the masses, the cut-off frequency comes out to be $\omega_c= 31.3\,$Trad/s. Since the frequency of the light irradiation $\left(\sim 10^{15}\,\text{Hz}\right)$ is about three orders of magnitude higher than the phonon vibrational frequency, we assume that the irradiation will not affect the lattice vibration significantly and hence  the effect of irradiation on lattice vibration can safely be ignored.

\begin{figure}[ht!]
\includegraphics[width=0.23\textwidth,height=0.15\textwidth]{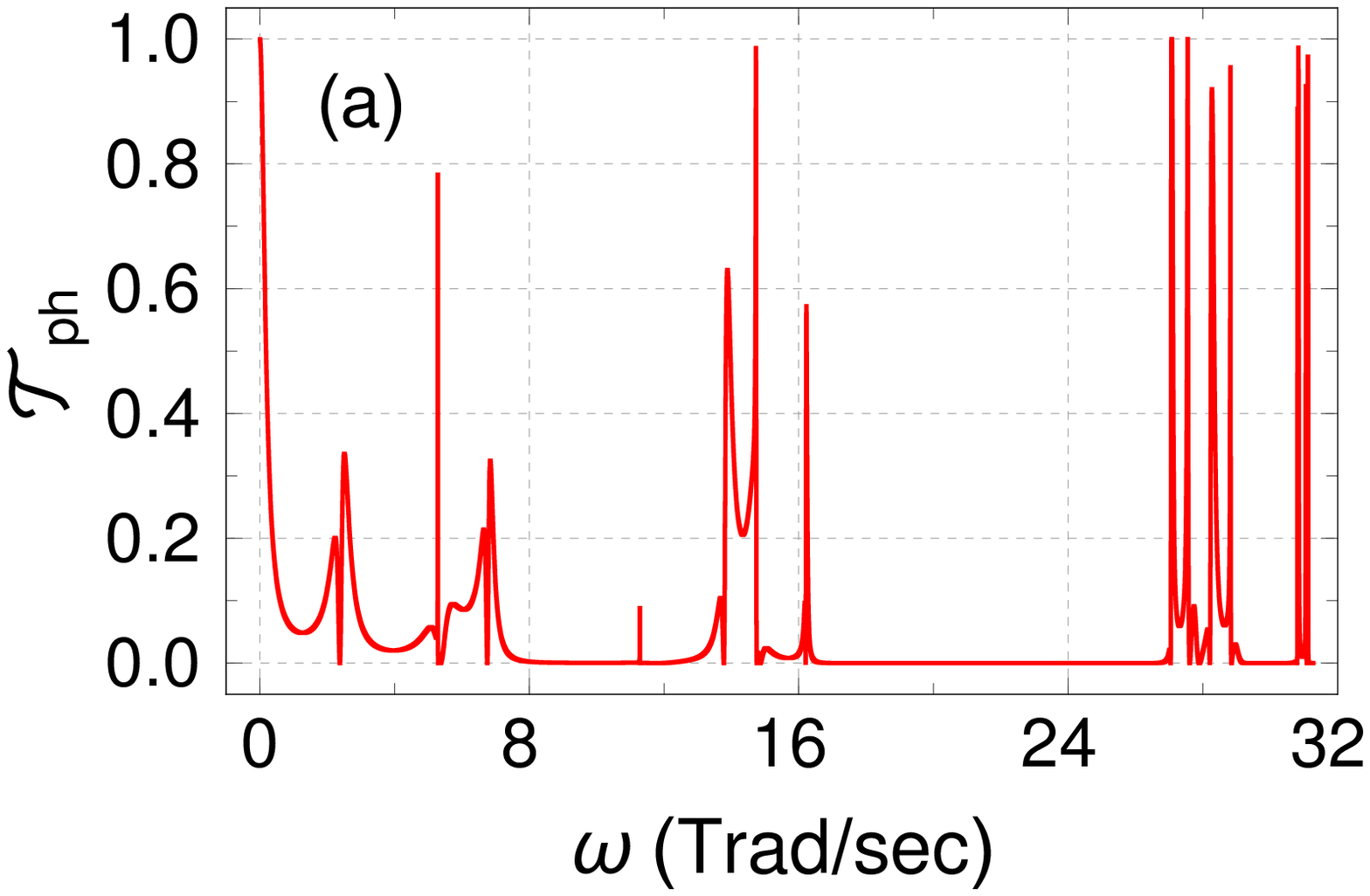} \hfill
\includegraphics[width=0.23\textwidth,height=0.15\textwidth]{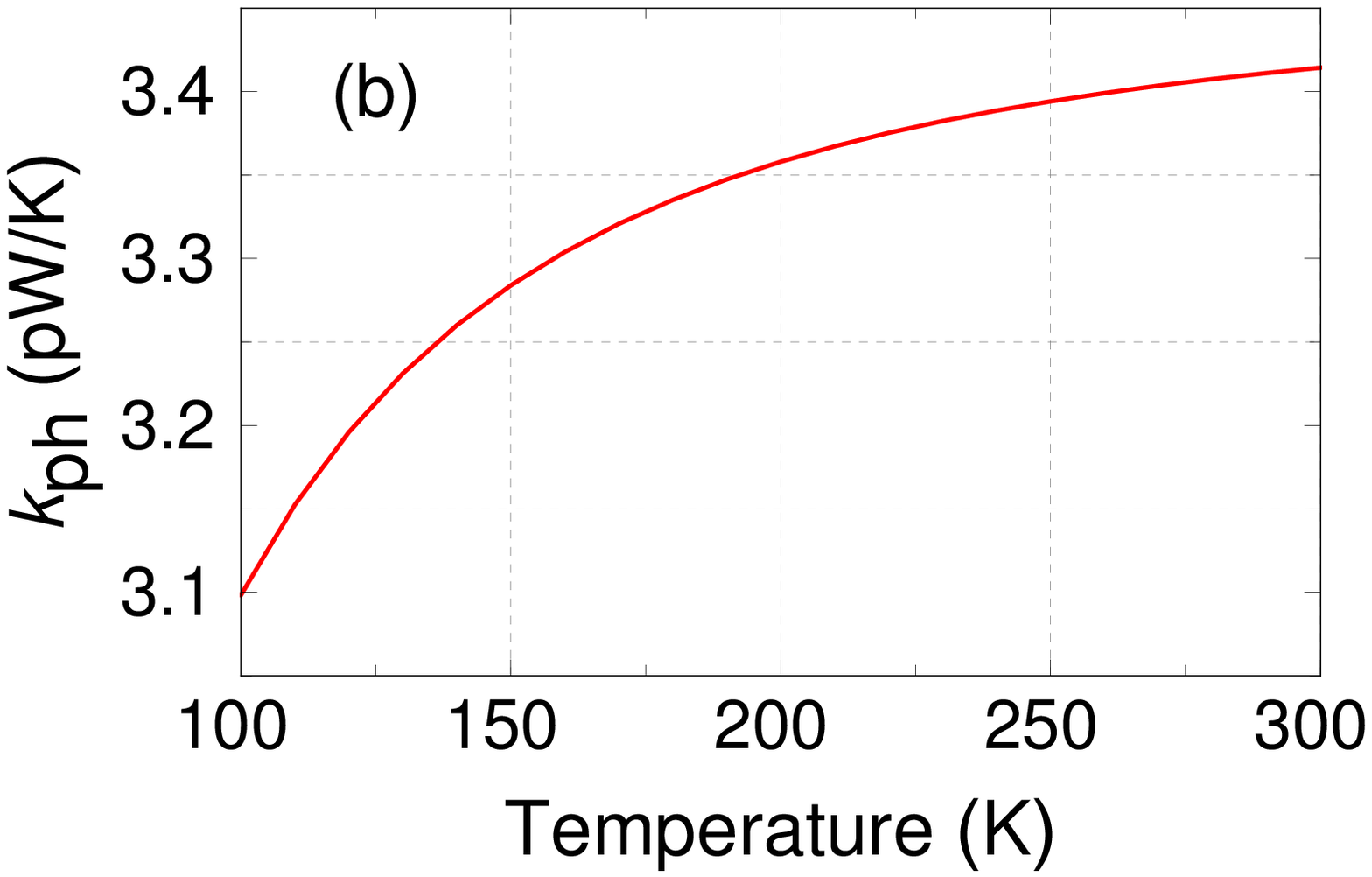}
\caption{(Color online). (a) Phonon transmission $\mathcal{T}_{ph}$ as a function of phonon angular frequency $\omega$. (b) Phonon thermal conductance $k_{ph}$ as a function of temperature $T$.}
\label{tph}
\end{figure}

Analogous to electronic transport, the localization phenomena are also expected to occur in the phonon transmission due to the fractal nature of SPG. In Fig.~\ref{tph}(a), the phonon transmission coefficient $\mathcal{T}_{ph}$ is plotted as a function of the phonon  frequency $\omega$ for a 5th generation SPG. Here the acoustic vibrational modes are greatly suppressed within the frequency range $\sim 16 - 27$ $\,$Trad/s. For 
the higher generation SPGs, we get more localized phonon modes which we confirm by studying the phonon transport up to the 8th generation SPG (not shown here). Such localization of phonon modes has also been studied in asymmetric harmonic chains~\cite{hopkins-prb} and observed in 2D fractal heterostructures~\cite{han-mrx,krishna}. The corresponding phonon thermal conductance $k_{ph}$ as a function of temperature is shown in Fig.~\ref{tph}(b). Here $k_{ph}$ is of the order of a few pW/K, about two orders of magnitude lower than its electronic counterpart. It increases systematically but very slowly with temperature.

\begin{figure}[ht!]
\begin{center}
\includegraphics[width=0.35\textwidth]{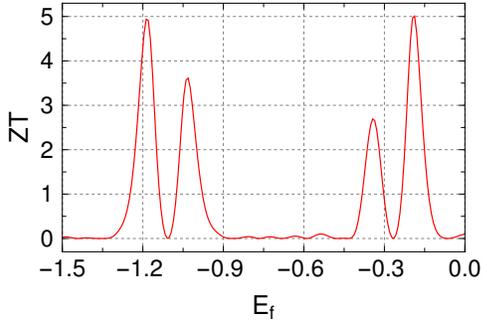}
\caption{(Color online). $ZT$ as a function of the Fermi energy $E_F$. The SPG configuration and all the other  parameters are identical as mentioned in Fig.~\ref{tgsk}.}
\label{zt}
\end{center}
\end{figure}

With all the TE quantities, we finally calculate the FOM. $ZT$ as a function of the Fermi energy is presented in Fig.~\ref{zt}. The maximum $ZT$ is $\sim 5$ around the mobility edge. This is, of course, a highly favorable response and a direct consequence of the existence of the asymmetric function across the mobility edges for the irradiated SPG network.

Here it is important to note that all the results studied in this communication are worked out for a set of typical parameter values. Now, if we had considered a real material like say Ge or Si, then the on-site energies would be non-zero and the NNH integrals would be of the order of eV~\cite{ge-hop} which we have considered in the present work.  Any non-zero on-site energy simply provides a shift of the energy spectrum, keeping all the physical pictures unaltered.

\section{Summary}

In conclusion, we have given a new prescription to get mobility edge separating the localized and extended states by irradiating an SPG network, and we exploit the existence of mobility edge in TE applications. Such an attempt is completely new to the best of our knowledge. The system under consideration has been described within a tight-binding framework, and the irradiation effect has been incorporated using the Floquet-Bloch ansatz in the minimal coupling scheme. The electronic and phononic transmission coefficients have been evaluated using the standard Green's function formalism based on Landauer-B\"{u}ttiker approach. At first, a higher generation SPG has been studied to observe the existence of mobility edge by studying the electronic transmission spectrum and energy eigenvalues in the presence of light irradiation. After that effect of light irradiation on the energy spectrum has been examined in detail. Finally, we have discussed the TE performance of a driven SPG by analyzing the different TE quantities such as the electrical conductance, thermopower, and thermal conductance due to electrons and phonons. Our essential findings are summarized as follows.\\
$\bullet$ Multiple mobility edges have been observed in the presence of light.\\
$\bullet$ The highly degenerate eigenvalues of SPG spread out in the presence of light.\\
$\bullet$ The bandwidth can be modified by tuning the light parameters.\\
$\bullet$ The fractal nature of an undriven SPG remains intact in the presence of circularly polarized light with a renormalized bandwidth.\\
$\bullet$ A band inversion has been observed in the presence of a circularly polarized light by tuning the field amplitude.\\
$\bullet$ Controlling the phase of the vector potential it is possible to have a highly conducting SPG from a poor one. This particular feature can be exploited to engineer switching devices. \\
$\bullet$ The electrical conductance and thermal conductance due to electrons are suppressed appreciably around the mobility edge while the thermopower acquires high value.\\
$\bullet$ The localization phenomenon has also been observed in the phonon transmission spectrum.\\
$\bullet$ The phonon thermal conductance is two orders of magnitude lower than its electronic counterpart.\\
$\bullet$ We have found that FOM is large and greater than unity by setting upon the Fermi energy around the mobility edge.

Our analysis can be utilized to investigate electronic and spin-dependent transport phenomena in similar kinds of fractal lattices and also to design fascinating spintronic and electronic devices.



\end{document}